\documentclass[floatfix,twocolumn,amsmath,amssymb,aps,prl,showpacs,letter,superscriptaddress]{revtex4}

\usepackage{amsmath}
\usepackage{amssymb}
\usepackage{graphicx}
\usepackage{subfigure}
\usepackage{epstopdf}
\usepackage{color}
\usepackage{ulem}
\usepackage{bm}

\usepackage{graphicx}
\usepackage{epsfig}
\usepackage{rotating}
\usepackage{amsmath}
\usepackage{amsfonts}
\usepackage{amssymb}
\usepackage{enumerate}
\usepackage{longtable}
\usepackage{dcolumn}% Align table columns on decimal point
\usepackage{bm}

\begin{document}

\title{Is the Yb$_2$Ti$_2$O$_7$ pyrochlore a quantum spin ice?
}

\author{R. Applegate}
\affiliation{Physics Department, University of California at Davis, Davis, CA 95616}
\author{N. R. Hayre}
\affiliation{Physics Department, University of California at Davis, Davis, CA 95616}
\author{R. R. P. Singh}
\affiliation{Physics Department, University of California at Davis, Davis, CA 95616}
\author{T. Lin}
\affiliation{Department of Physics and Astronomy, University of Waterloo, Waterloo,
Ontario, N2L 3G1, Canada}
\author{A. G. R. Day}
\affiliation{Department of Physics and Astronomy, University of Waterloo, Waterloo,
Ontario, N2L 3G1, Canada}
\affiliation{D\'epartement de Physique, Universit\'e de Sherbrooke, Sherbrooke, Qu\'ebec, J1L 2R1, Canada}
\author{M. J. P. Gingras}
\affiliation{Physics Department, University of California at Davis, Davis, CA 95616}
\affiliation{Department of Physics and Astronomy, University of Waterloo, Waterloo,
Ontario, N2L 3G1, Canada}
\affiliation{Canadian Institute for Advanced
Research, 180 Dundas St. W., Toronto, Ontario, M5G 1Z8, Canada}

\date{\rm\today}

\begin{abstract}
We use numerical linked cluster (NLC) expansions to compute the specific heat, $C(T)$, and entropy, $S(T)$,
of a quantum spin ice model of Yb$_2$Ti$_2$O$_7$
using anisotropic exchange interactions
recently determined from inelastic neutron
scattering measurements and find
good agreement with experimental calorimetric data. In the perturbative weak quantum regime, this
model has a ferrimagnetic ordered ground state, with two peaks in $C(T)$:
a Schottky anomaly signalling the paramagnetic
to spin ice crossover followed at lower temperature by 
a sharp peak accompanying 
a first order phase transition to the ferrimagnetic state. 
We suggest that the two $C(T)$ features observed in
Yb$_2$Ti$_2$O$_7$ are associated with the same physics.
Spin excitations in this regime consist of weakly confined spinon-antispinon pairs.
We suggest that conventional ground
state with exotic quantum dynamics will prove
a prevalent characteristic of many
real quantum spin ice materials.
\end{abstract}


\maketitle

%%%%%%%%%%%%%%%%%%%%%%%%%%%%%%%%%%%%%%%%%%%%%%%%%%%%%%%%%%%%%%%%%%

%{\it Paragraph 1: Why interested in general in quantum mechanics/quantum fluctuations in condensed matter
%physics, spin liquids. Can be shortened. Could start at `Over the past twenty years'}

%If sufficiently large, quantum mechanical zero-point fluctuations can inhibit the development of
%conventional long-range order in condensed matter systems and allow for the manifestation of
%fascinating phenomena.
%A prominent example is the failure of helium to crystallize at atmospheric pressure down to the lowest
%accessible temperature, which hence allow for superfluidity, a macroscopic quantum coherent state of matter. 

The experimental search for quantum spin
liquids (QSLs),  magnetic systems disordered by large quantum fluctuations, 
has remained unabated for over twenty years \cite{Balents_nature}.
One direction that is rapidly gathering momentum is the search for QSLs among materials 
that  are close relatives to spin ice systems \cite{spin_ice_review}, 
but with additional quantum fluctuations, or {\it quantum spin ice} \cite{molavian,onoda}.

%{\it Paragraph 2: What are spin ices?

Spin ices are found among insulating pyrochlore oxides, such as R$_2$M$_2$O$_7$ 
(R=Ho, Dy; M=Ti, Sn)  \cite{GGG_RMP}.
In these compounds, the magnetic R rare earth ions sit on a 
%reside on the sites of a 
lattice of corner-sharing tetrahedra, experiencing
a large single-ion 
%crystal-field 
anisotropy forcing the magnetic moment
to point strictly ``in'' or ``out'' of the two tetrahedra it joins  (see. Fig. 1a). 
Consequently, the direction of a moment can be described by a classical Ising spin ~\cite{spin_ice_review}.
%variable ~\cite{spin_ice_review}. 
In these materials, the combination of nearest-neighbor exchange and long-range magnetostatic
dipolar interactions lead to an exponentially large number of low-energy states characterized by 
two spins pointing in and two spins pointing out on each tetrahedron (see Fig. 1a).
This energetic constraint is equivalent to the Bernal-Fowler ice rule 
which gives water ice a residual entropy  $ S_{\rm P} \sim {k_{\rm B}}(\frac{1}{2})\ln(3/2)$ per proton,
 estimated by Pauling \cite{Pauling} and in good agreement with experiments on water ice \cite{Giauque}.  
Since they share the same ``ice-rule'', the (Ho,Dy)$_2$(Ti,Sn)$_2$O$_7$ pyrochlores also possess
a residual low-temperature Pauling entropy $S_{\rm P}$ \cite{entropy}, hence the name  spin ice.
The spin ice state is not thermodynamically distinct from the paramagnetic phase. 
Yet,  because of the ice-rules, it is a strongly correlated state of matter --
 a {\it classical} spin liquid of sorts \cite{Balents_nature,spin_ice_review}.

\begin{figure}
\begin{center}
\includegraphics[width=8cm]{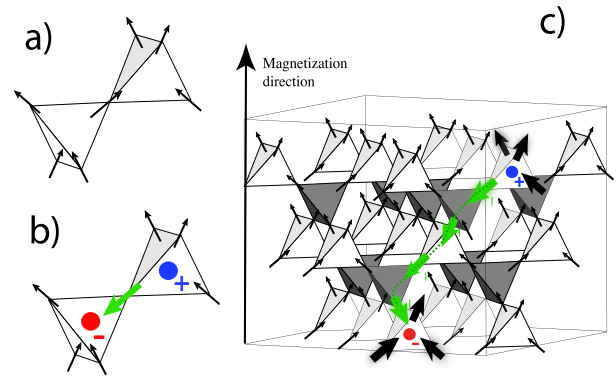}
\caption{\label{fig1}
(a) Two neighboring tetrahedra with spins in their two-in/two-out ground state, (b) 
spinon/antispinon pair, 
%Monopole-antimonopole pair, 
(c) spinon/antispinon pair 
%Monopole-antimonopole pairs 
separated by a (green) string of misaligned spins  in the pyrochlore lattice.
}
\end{center}
\end{figure}

%{\it Paragraph 3; Enters QSI}

For infinite Ising anisotropy,  quantum 
% fluctuation 
effects are absent ~\cite{spin_ice_review}.
However, these can be restored when considering the realistic situation of {\it finite} anisotropy.
In two  closely related papers, 
Hermele {\it et al.}~\cite{Hermele} and Castro-Neto {\it et al.}~\cite{Castro-Neto} considered
effective spins one-half on a pyrochlore lattice where
the highly degenerate classical spin ice state is promoted 
via quantum fluctuations to a QSL with fascinating properties.
This QSL is described by a compact lattice quantum electrodynamics (QED) -like theory.
In this QSL state inherited from the parent classical spin ice, 
the ice-rules amount to  a divergence-free coarse-grained fictitious
electric field whose sources are deconfined spinons while the sources of the 
canonically conjugate field are deconfined monopoles \cite{monopole_vs_spinon}, 
along with a gauge boson (``artificial photon'').
%both  defects accompanied by a gauge boson (``artificial photon'').
%inherited by the underlying U(1) symmetry of the theory. 

%{\it Paragraph 4; QSI in a material context and the promise of Yb2Ti2O7}

Recent numerical studies have found evidence that QED-like phenomena may be 
at play in some minimal  quantum spin ice (QSI) lattice models~\cite{QED_numerics}  -- but does the
QSI picture apply to real materials?
Also, should a QSI
%{\it quantum spin ice}  
state  be solely defined by whether or not a QSL state is realized?
While a QSI picture has been suggested relevant to 
%has been proposed to be relevant for the 
the QSL behavior in Tb$_2$Ti$_2$O$_7$ \cite{molavian} and Pr$_2$M$_2$O$_7$ \cite{onoda},
% for some time now,
intense experimental \cite{hodges,ross_prl,cao,thompson_prl,ross,yaouanc_prb,ross_prb,chang}
and theoretical \cite{cao,thompson_prl,ross,chang,malkin,onoda_conf,thompson_jpcm,savary,oleg} 
interest has recently turned to Yb$_2$Ti$_2$O$_7$ (YbTO), which
has been argued to be on the verge of realizing a QSL originating from QSI physics.
%has been argued to be either a QSL or, at least, to be on the verge of being one.
In fact, the combination of (i) an unexplained transition at $T_c\sim$ 0.24 K \cite{hodges,blote}, 
(ii)  the controversial
evidence for 
%conventional 
long-range order below $T_c$ ~\cite{yasui,gardner_YbTO} and (iii) the high sensitivity 
of the low-temperature ($T<300$ mK) behavior to sample preparation conditions \cite{yaouanc_prb,ross_prb}
are all tantalizing evidence that YbTO has a 
%very 
fragile and perhaps 
%highly 
unconventional ground state.
%phase. 
Thus, explaining YbTO is a key
%important 
milestone in the study of QSI in a materials context.

In this paper, we first use the numerical linked cluster (NLC) method \cite{series,rigol} to 
calculate the heat capacity, $C(T)$, and entropy, $S(T)$, 
of a microscopic model for YbTO with exchange parameters, $\{ J_e \}$, taken from Ref.~\cite{ross}.
This calculation, which converges down to about 1 K, agrees well
with experiments. It demonstrates that YbTO is indeed a spin-half, anisotropic
exchange model, with $\{ J_e \}$ determined from magnon energies in the 
strong-field polarized paramagnet regime \cite{ross}.
Our work suggests that a two-peaked $C(T)$ structure is natural in YbTO and 
should be present in the best (``quality'') samples \cite{yaouanc_prb,ross_prb}.
Below the higher temperature $C(T)$ hump near 2 K, the system has a residual $S(T)$ 
comparable to $S_{\rm P}$, but without a clean $S(T)\approx S_{\rm P}$ plateau developing upon cooling. 
We propose that the lower temperature sharp peak in $C(T)$ is associated
with a strongly first order transition to a 
ferrimagnetic state. 
Such a behavior is indeed found in our study when the quantum (non-Ising) exchanges are small.
Finally, we argue  that 
despite a conventional 
%semi-classical ferrimagnetic 
ground state, 
the spin excitations consist of spinon/antispinon pairs
connected with (Dirac-like \cite{castelnovo}) strings of reversed spins, 
whose confinement length $l_s$ diverges in the
% perturbative 
limit of  small  quantum exchanges. 
We propose that these excitations should ultimately form 
the basis for describing
what we expect to be highly unconventional inelastic  
neutron spectra ~\cite{oleg}.

{\it Model \& Method} --
The anisotropic exchange QSI model is defined by the nearest-neighbor Hamiltonian \cite{ross,savary}
on the pyrochlore lattice 
%The model is defined by a nearest-neighbor ansotropic exchange QSI Hamiltonian, 
%${\cal H}_{\rm QSO}$, on the pyrochlore lattice
\begin{eqnarray}
{\cal H}_{\rm QSI}
&=&\sum_{<i,j>} \{ J_{zz}S_i^z S_j^z -\lambda J_\pm (S_i^+ S_j^- +S_i^-S_j^+)\nonumber \\
        & &+ \lambda J_{\pm\pm} [\gamma_{ij}S_i^+S_j^+ + \gamma_{ij}^* S_i^-S_j^-]\nonumber \\
        & &+ \lambda J_{z\pm} [(S_i^z(\zeta_{ij}S_j^+ +\zeta_{i,j}^*S_j^-) + i \leftrightarrow j] \} .
\label{Hqsi}
\end{eqnarray}
$\gamma_{ij}$  is a $4 \times 4$ complex unimodular matrix, and $\zeta=-\gamma^*$ \cite{ross}.
The $\hat z$ quantization axis is along the local $[111]$ direction,
and $\pm$ refers to the two orthogonal local directions. We 
%will 
take $\lambda=1$, except when stated otherwise.

Recently Ross {\it et al.} \cite{ross} used inelastic neutron scattering data in high fields to deduce the
 $\{ J_{\rm e} \}$ exchange parameters for YbTO: 
 $J_{zz}=0.166\pm 0.04$, $J_\pm=0.05\pm 0.01$, $J_{\pm\pm}=0.05\pm 0.01$, 
and $J_{z\pm}=-0.14\pm 0.01$, all in meV.
These parameters have also been determined through an analysis of the zero-field
energy-integrated paramagnetic neutron scattering \cite{thompson_prl,chang}, but the values of the $\{ J_e \}$ parameters
disagree significantly -- an issue that we address in the supplementary material \cite{see_sup_mat}.

%Numerical linked cluster (NLC) 

NLC expansions provide a controlled way of calculating macroscopic
properties of a thermodynamic system \cite{series,rigol}.
 By summing up contributions from clusters upto some
size, one can obtain properties in the thermodynamic limit, which include all terms in high 
temperature expansions upto some order. Furthermore, since the contributions of the
clusters are entirely included for all temperatures, all short distance physics is 
fully incorporated, and thus can converge down to lower temperatures than
a (high-temperature, $T$) series expansion \cite{series} in $1/T$. 
NLC is particularly suited to the study of spin ice systems. 
It was recently shown that for classical spin ice models, 
just first order NLC based on a single tetrahedron, gives 
$C(T)$ and $S(T)$ for all $T$ within a few percent accuracy \cite{oitmaa}.

Here, we calculate the thermodynamic properties of the exchange QSI model of Eq.~(\ref{Hqsi})
using tetrahedra-based NLC 
%. Results are obtained 
upto 4$^{\rm th}$ order \cite{see_sup_mat}. Euler extrapolations \cite{Euler_method} 
are used to eliminate some alternating pieces in the expansion, which further
improves the convergence of the calculations to lower $T$. 
In zero field, there is only one cluster in each of the first three orders, and three clusters
in the fourth order \cite{see_sup_mat}. 
The different $g$-tensor elements on different sites (expressed in a global frame) \cite{thompson_jpcm}
mean that many more clusters are needed for calculating 
field-dependent $C(T)$, magnetization and
susceptibility, and these will be presented elsewhere.

Figure 2 shows  $C(T)$ calculated with different NLC orders.
By 4$^{\rm th}$ order, there is good convergence to temperatures below
the $C(T)$ peak at $\sim 2$ K.
Applying Euler transformations \cite{Euler_method} improves the convergence down to slightly below 1 K.
The experimental data from Refs. ~\cite{blote}, shown for comparison,
agree well with the NLC results. Here, we used
the mean values of the $\{ J_{\rm e} \}$ from Ref.~\cite{ross} and did not adjust any parameters.
Given the variability in the experimental $C(T)$ data from one  group to another \cite{yaouanc_prb,ross_prb,chang,see_sup_mat},
it does not seem useful at this time to search for
$\{ J_{\rm e} \}$  parameters giving a better fit. This agreement shows that the $\{ J_e \}$
%exchange
 parameters are not substantially renormalized compared to the high (5 Tesla) field values \cite{ross}.
Using the  $\{ J_{\rm e} \}$ 
%exchange constants 
of Refs.~\cite{thompson_prl,chang} gives substantially different $C(T)$  results \cite{see_sup_mat}.

\begin{figure}[ht]
\begin{center}
\includegraphics[angle=0,width=9cm]{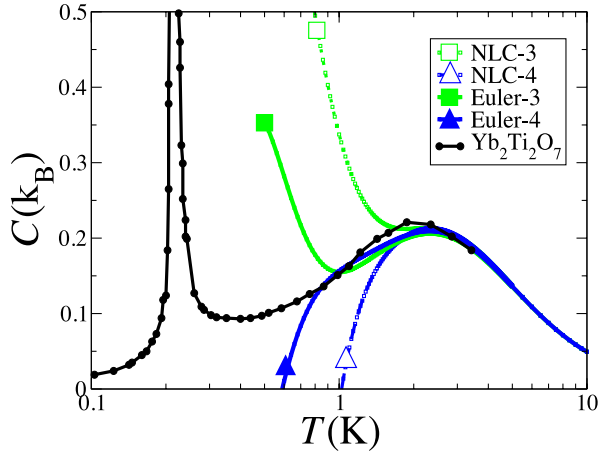}
\caption{\label{fig2}
Specific heat, $C(T)$, per mole of Yb for the model parameters in Ref.~\cite{ross}, 
in units of the Boltzmann constant $k_{\rm B}$, calculated via NLC
(up to 4$^{\rm th}$ order NLC together with Euler extrapolations) are compared with experimental data
for Yb$_2$Ti$2$O$_7$. 
%Several sets of experimental data are shown.
The black circles are data from Ref.~\cite{blote}.
%The open violet diamonds and violet circles are the data for the WHAT SAMPLES
%of Ref.\cite{Yaouanc}
}
\end{center}
\end{figure}

\begin{figure}[h]
\begin{center}
\includegraphics[angle=0,width=9cm]{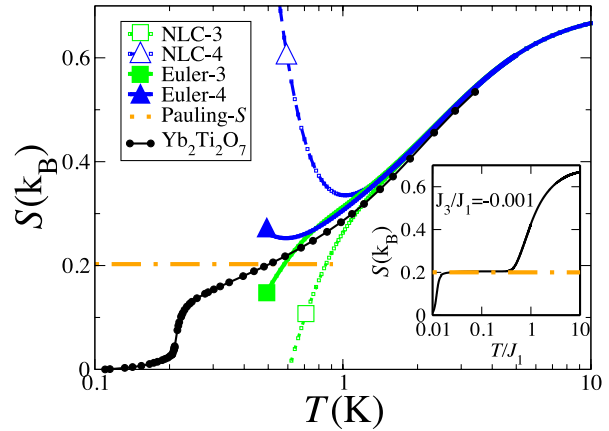}
\caption{\label{fig3}
Entropy, $S(T)$, per mole of Yb, in units  $k_{\rm  B}$ following the methods
described in the caption of Fig. 2.
The black circles are obtained by integrating the data from Ref.~\cite{blote} excluding
the nuclear (hyperfine) contribution.
The Pauling entropy $S_{\rm P} \sim { k_{\rm B}\over 2 }\ln{3\over 2}$ is shown
as a horizontal line.
The inset shows $S(T)$ in the perturbative 
regime with $J_3/J_{zz}=-0.001$. A clear plateau at $S(T) \approx S_{\rm P}$ 
is seen, followed at lower $T$ by a precipitous drop of $S(T)$ (i.e.
latent heat) accompanying the transition to long range FM order \cite{see_sup_mat}.
}
\end{center}
\end{figure}

Figure 3 shows $S(T)$ calculated by NLC, together with the entropy
obtained by integrating $C(T)/T$ data of Ref.~\cite{blote}.
We found the data from Ref.~\cite{blote} ideally suited to perform this comparison \cite{see_sup_mat}.
The entropy converges to lower temperature slightly  better than $C(T)$ where, 
with Euler transformations, $S(T)$ converges down to about 0.7 K, matching
well with the experimental entropy values over the overlapping temperature range.

{\it Perturbative considerations} --
% for ${\cal H}_{\rm QSI}$} --
In order to better understand the properties of this system, we
turn to the perturbative regime $\lambda\ll 1$ in Eq.~1~\cite{ross,savary}. 
%where all ``quantum'' 
%$J_e \ne J_{zz}$ exchange constants are small compared to $J_{zz}$ ~\cite{ross,savary}. 
To second order in $\lambda$, only $J_{z\pm}$, by far the largest quantum term for YbTO,
leads to a degeneracy-lifting classical potential
 for different spin-ice configurations. 
It amounts to a 
fluctuation-induced ferromagnetic exchange constant 
$ J_3\equiv -3\lambda^2 J_{z\pm}^2/J_{zz}$ \cite{savary}  between shortest distance
spins on the same tetrahedral sublattice that share a neighbor \cite{on_3rd_neighbors}.
It leads to the selection of a ${\bm q}=0$ long-range ordered 
ground state in which all tetrahedra
are in the same configuration and the spins develop a small ferromagnetic
moment along one of the $\langle 1 0 0 \rangle$ cubic directions.
This ${\bm q}=0$ ferrimagnet (FM) lacks 
%any of 
the Coulombic physics originally present in the $J_{zz}$-only
spin ice model \cite{henley_annual_rev_cmp}.

To calculate $C(T)$ and $S(T)$ in the perturbative regime at low $T$,
we turn to classical loop Monte Carlo simulations  ~\cite{loop_MC} of the $J_3-J_{zz}$ model \cite{see_sup_mat}.
%The Monte Carlo results agree with NLC for the high temperature peak and
These reveal a very sharp lower temperature peak signalling a first order 
phase transition to a  ${\bm q}=0$ state (see Fig. S5  \cite{see_sup_mat}).
%correlations in the ground state.

{\it Excited states in the perturbative regime: spinons and strings} --
A surprise of the perturbative treatment is that, while
the ground state is 
%entirely 
classical, the spin-flip  
excitations remain 
%highly 
non-trivial and of quantum nature.
This is because, once a spin is flipped
in a spin-ice state, creating a  spinon/antispinon pair \cite{monopole_vs_spinon},
the pair can  hop through $J_{z\pm}$ acting through {\it first order} degenerate perturbation theory.
Thus, the dispersion in the excited state  manifold is $\lambda J_{z\pm}$, {\it much larger}
than the dispersion 
within the low-energy manifold of spin ice states,
which is only $\lambda^2 J_{z\pm}^2/J_{zz}$. 
%\ll \lambda J_{z\pm}$.

A sketch of a spinon/antispinon pair
is shown in Fig. 1b and 1c.
Note that only spins inside the tetrahedron ``already'' containing spinons 
are flippable in first order degenerate perturbation theory.
%and can propagate (hop) the defect.
Hence, the connecting string of misaligned spins 
can only fluctuate by higher order  processes involving
closed loops with alternating in-out spins \cite{oleg}.
Thus the renormalized string tension per unit length remains finite and of order $J_3$. 
One can estimate the typical string length as the length, $l_{\rm s}$,
at which the cost of the string becomes comparable to the delocalization energy
of the spinon/antispinon pair. 
The string energy per unit length goes as $\sim J_3 \sim \lambda^2$, whereas 
the delocalization energy (spinon bandwidth) goes as $\lambda$. 
This leads to $l_s$ scaling as 
%$l_{\rm s} \sim $
$1/\lambda$, which diverges as $\lambda \rightarrow 0$.
%perturbative regime, signalling a return to the classical spin
%ice with deconfined excitations \cite{castelnovo}.

%Thus, perturbation theory leads to a very rich picture at $T=0$,
%where the ground state is conventional but the spin excitations remain
%highly non-trivial. 
A detailed theory of neutron scattering in
this ferrimagnetic phase is not attempted here, but we anticipate it to
%closely 
follow the 
%recent 
proposal of Ref.~\cite{oleg}.
At temperatures above the transition to the ${\bm q}=0$
long-range ordered state, 
the system explores the classical two-in/two-out spin ice states and 
should display singularities (pinch points, PPs) in neutron scattering \cite{henley_annual_rev_cmp}  
rounded off by the finite density of
 thermally excited spinon/antispinon defects \cite{monopole_vs_spinon,henley_annual_rev_cmp}.
While the system has thermally smeared PPs above the ferrimagnetic
transition and no static PPs well below the transition, it may
display some remnant of PPs in the spin dynamics at higher energies.
%associated with  the diverging string length. 
These interesting issues deserve further  attention.

\begin{figure}
\begin{center}
\includegraphics[angle=0,width=6cm]{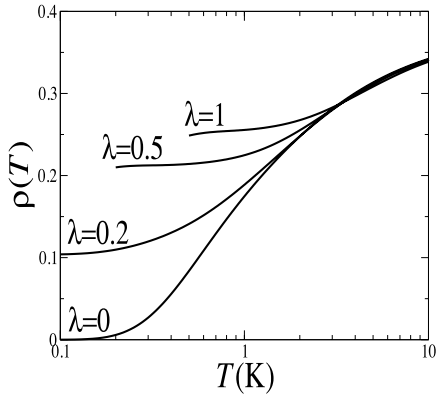}
\caption{\label{fig4}
Monopole defect density, $\rho(T)$, calculated using NLC,
shown down to a temperature where
$3$rd and $4$th order Euler Transforms agree.
Here, quantum exchanges are scaled with respect to YbTO parameters by
different values of $\lambda$.
}
\end{center}
\end{figure}

{\it Beyond the $\lambda \ll 1$  regime} --
Why is the transition temperature of YbTO so low?
As discussed by Ross {\it et al.}~\cite{ross},
the low $T$ peak in $C(T)$ is at a temperature
lower than mean-field theory by an order of magnitude.
Comparing $C(T)$ for the quantum model with different $\lambda$ with
the corresponding classical model with the perturbative $J_3/J_{zz}$ value 
%gives us
provides a 
% some 
hint of the reason why \cite{see_sup_mat}. 
It shows that, in the classical model, 
the long-range order keeps steadily moving up with increased $J_3$, even beyond the short-range
order $C(T)$ peak. In contrast, the quantum systems, with different $\lambda$
continue to display
%have 
a short-range order $C(T)$ peak and presumably long-range order only
occurs at a much lower $T$.
Perturbative considerations here have an analogy
with strong coupling studies of Mott physics in the Hubbard model, where
the N\'eel temperature
%In a 3D Hubbard model, it is well known that the N\'eel temperature
first increases with $t$ as $t^2/U$ but then begins decreasing when
the system moves away from the 
 perturbative small $t/U$ regime.
We propose that a similar non-monotonic $T_c$ 
arises in this QSI model due to enhanced quantum fluctuations.

Another argument for a reduced $T_c$ comes from considering
the temperature dependence of the defect (spinon/antispinon) 
monopole density, $\rho(T)$, as calculated by NLC (see Fig.~4 and Figs. S3 and S4 \cite{see_sup_mat}).
To illustrate the point, we show the behavior for several different
$\lambda$ values.
Convergence increases to lower $T$, with decreasing $\lambda$, as expected.
One finds that as $T$ drops 
%system goes 
below  the hump in $C(T)$, 
$\rho(T)$ displays a plateau-like region, whose value
increases steadily with increasing $\lambda$.
This indicates that the states {\it within} the 
%(effective \& renormalized, hence quantum) 
spin-ice manifold develop large 
spinon/antispinon spectral weight,
%textures, 
thus strongly renormalizing all low energy scales 
%fields and effective interactions. 
and, presumably, leading to 
%his should lead to a 
reduced $T_c$.

{\it Discussion: What constitutes an exchange QSI?} --
We suggest that  a double-peaked $C(T)$ with an entropy
between the peaks comparable to $S_{\rm P}$ is
the hallmark of an exchange quantum spin ice (QSI). 
However, one is unlikely to find an exact plateau at $S(T) \approx S_{\rm P}$
outside the perturbative (small $\lambda$) regime. 
Such a double-peaked structure and 
quasi-separation of the energy/temperature scales associated with
short and long-range physics has also been suggested
for other systems where quantum spin liquid physics may apply \cite{kagome}.
%This has long been sought as a signature of
%Resonating Valence Bond physics in kagome systems, but has proved 
%elusive there.\cite{kagome}

According to the gauge mean-field theory of Ref.~\cite{savary},
at low temperature below which 
short-range spin ice correlations develop, a system may
exhibit either a conventional ferrimagnetic (FM) order, 
a Coulombic ferromagnet (CFM) or a full-blown quantum spin-liquid (QSL), depending on 
its  quantum exchange parameters.
%of the system. 
The largest quantum exchange terms in YbTO is $J_{z\pm}$,
which favors the FM state, which we believe is the origin
of the 0.24 K transition in the best samples \cite{yasui}.
 It remains to be seen if there are real
materials for which $J_{\pm}$, which favors the QSL \cite{Hermele,Castro-Neto,savary}, 
is the dominant quantum term.
Nevertheless, even when the ground state is FM, the
excitations remain highly exotic, consisting 
% They consist 
of spinon-antispinon pairs separated by long strings. This non-trivial feature is derived from
the underlying spin-ice physics.
%and thus qualifies all such systems
%as quantum spin ices (QSI).
Finally, as one notes that $J_{z\pm}$ is strictly zero for
non-Kramers ions (e.g. Pr, Tb) and that 
virtual crystal field excitations \cite{molavian} in Tb-based pyrochlores 
are a fundamentally different pathway 
from anisotropic superexchange \cite{onoda} to generate anisotropic $\{J_e\}$
couplings between effective spins one-half \cite{molavian,onoda}, 
the prospect to ultimately find a QSI-based QSL 
among rare-earth pyrochlores \cite{GGG_RMP} is perhaps promising.

\begin{acknowledgements}
This work is supported in part by NSF grant number  DMR-1004231, the NSERC of Canada
and the Canada Research Chair program (M.G., Tier 1).
We acknowledge very useful discussions with B. Javanparast, K. Ross and J. Thompson.
We thank P. Dalmas de R\'eotier for providing specific heat data 
of Ref.~[\onlinecite{yaouanc_prb}].
\end{acknowledgements}

%%%%%%%%%%%%%%%%%%%%%%%%%%%%%%%%%%%%%%%%%%%%%%%%%%%%%%%%%%%%%%%%%%

\section*{Supplementary Material}

This supplement provides the reader with further material to assist with some of the
technical materials of the main part paper

%%%%%%%%%%%%%%%%%%%%%%%%%%%%%%%%%%%%%%%%%%%%%%%%%%%%%%%%%%%%%%%%%%%%%%%%%%%%%%%%%%%%%%%%%%

\subsection*{Numerical Linked Cluster Method}

For the proposed QSI Hamiltonian ~\cite{ross},  the numerical linked cluster 
(NLC) method ~\cite{series,rigol} gives reliable quantitative properties of the system
in the thermodynamic limit
down to some temperature  by developing an expansion 
in connected tetrahedra that embed in the pyrochlore lattice. For each 
cluster, we perform an exact diagonalization (ED) and calculate physical 
quantities from the resulting spectrum and states. Once a property is 
calculated, the properties of all subclusters are subtracted to get the 
weight of the cluster $c$ denoted as $W(c)$. 
In the thermodynamic limit, an extensive property, $P$
is expressed as
\begin{equation}
P/N=\sum_c L(c) \times W(c),
\end{equation}
where $L_c$ is the count of the cluster, per lattice site.

We consider all clusters up to four tetrahedra, the largest diagonalization 
being a 13-site system. All states are required to calculate the partition 
function and thermodynamic quantities presented below.
The particular clusters to fourth order in our 
expansion are shown in Figure S1. 
%Note that sites in a single tetrahedron 
%pick up particular phases in our hamiltonian and that leads 
%to a distinction between clusters ABA and ABC.

%%%%%%%%%%%%%%%%%%%%%%%%%%%%%%%%%%%%%%%%%%%%%%%%%%%%%%%%%%%%%%%%%%%%%%%%%%%%%%%%%%%%%%%%%%

\subsection*{Computational Requirements}

NLC using the tetrahedral basis requires exact diagonalization of
increasingly large tetrahedral clusters.  Using modern hardware and
freely-available linear algebra routines,  diagonalizations for clusters
of one tetrahedron (four sites) and two tetrahedra (seven sites) could be
done in less than a second, while the three-tetrahedron (10-site) cluster
still required less than 10 seconds.  Computing only the spectrum for a single
four-tetrahedron (13-site) cluster required about 1200 seconds and more
than 1 GB of memory, while generating the full set of eigenstates required 
approximately 8 GB of memory. Note that the Hamiltonian of an N-site
cluster is a $2^N\times 2^N$ complex Hermitian matrix.
Exact diagonalizations of larger systems
are, in practice, limited by memory requirements. The next order calculation
will have $3$ more sites and the memory requirement will grow by a factor
of $64$.

%%%%%%%%%%%%%%%%%%%%%%%%%%%%%%%%%%%%%%%%%%%%%%%%%%%%%%%%%%%%%%%%%%%%%%%%%%%%%%%%%%%%%%%%%%

\subsection*{Euler Summation}

NLC generates a sequence of property estimates $\{P_n\}$ with 
increasing order $n$, where $P_n = \sum_{i=1}^n S_i$
and $S_i$ is some physical quantity calculated at the $i$th order.  
When such a sequence is found to alternate, its convergence 
can be improved by Euler Transformation \cite{Euler_method}. In general, given 
alternating terms $S_i = (-1)^i u_i$, the Euler Transform method
amounts to estimates,
\begin{equation}
u_0 - u_1 + u_2 - \ldots - u_{n-1} + \sum_{s=0} \frac{(-1)^s}{2^{s+1}}[\Delta^s 
u_n],
\end{equation}
where $\Delta$ is the forward difference operator
\begin{eqnarray}
 \Delta^0 u_n & = & u_n, \nonumber \\
 \Delta^1 u_n & = & u_{n+1} - u_n, \nonumber \\
 \Delta^2 u_n & = & u_{n+2} - 2 u_{n+1} + u_n, \nonumber \\
 \Delta^3 u_n & = & u_{n+3} - 3 u_{n+2} + 3 u_{n+1} - u_n, \ldots .
\end{eqnarray}
Usually, a small number of terms are computed directly, and the Euler 
transformation is applied to rest of the series.  In our case, where direct 
terms are available to fourth order, we begin the Euler
transform after the second order, 
so that the third and fourth order Euler-transformed property estimates are
\begin{eqnarray}
 P_{3,\text{E}} & = & S_0 + S_1 + S_2 + \frac{1}{2}S_3, \nonumber \\
 P_{4,\text{E}} & = & P_{3,\text{E}} + \frac{S_3+S_4}{4}.
\end{eqnarray}

\begin{figure}
\centering
\includegraphics[width=3.25in]{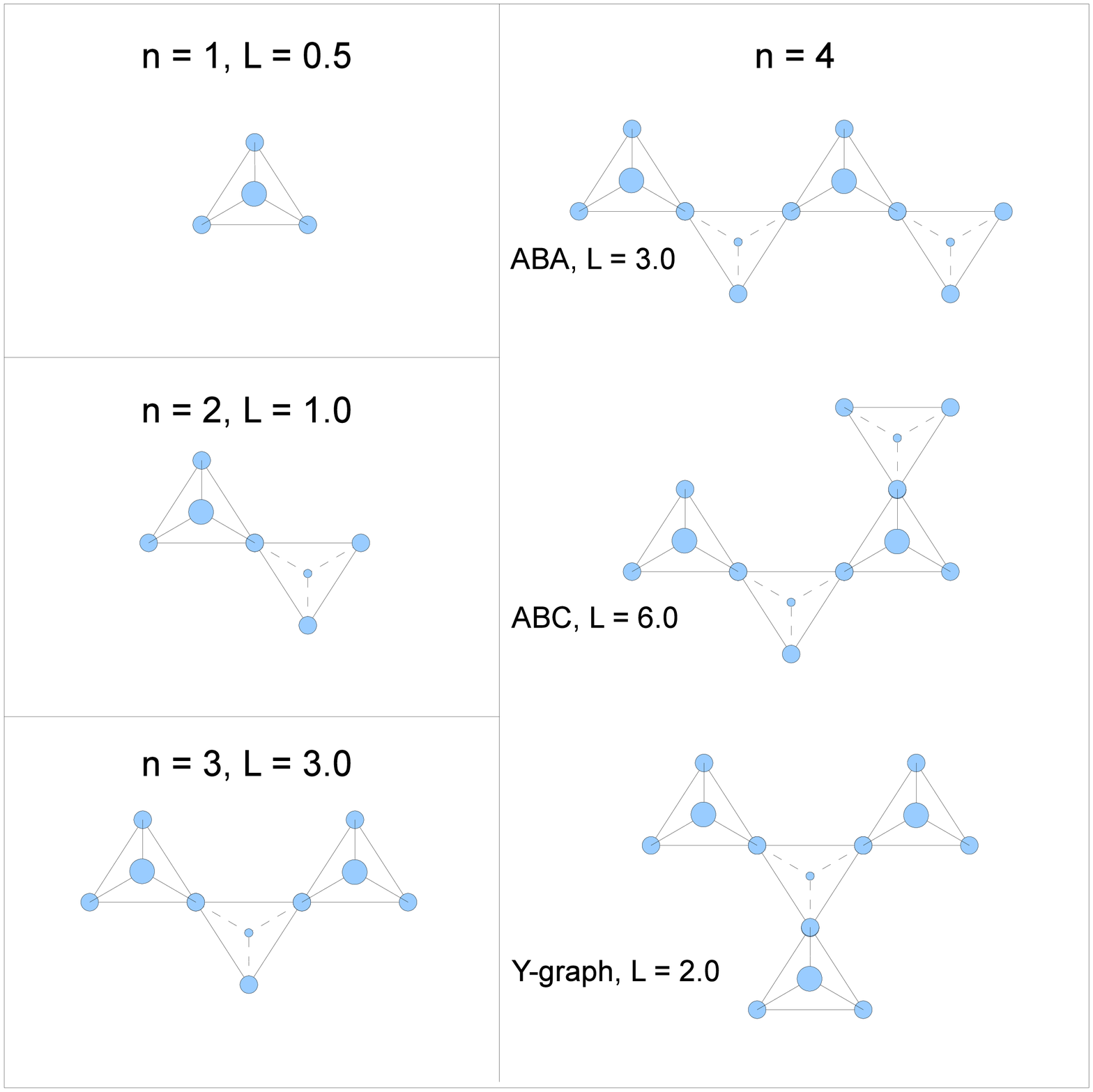}
\caption{S1: Clusters used for the zero-field NLC expansion in the tetrahedral 
basis, up to fourth order.  Each graph is accompanied by its lattice 
constant $L$. }
\label{NLC_graphs}
\end{figure}

%%%%%%%%%%%%%%%%%%%%%%%%%%%%%%%%%%%%%%%%%%%%%%%%%%%%%%%%%%%%%%%%%%%%%%%%%%%%%%%%%%%%%%%%%%

\subsection*{Various Hamiltonians and perturbative limit}

We use the notation of Ross {\it et al.} ~\cite{ross} and define
the quantum spin ice Hamiltonian as
\begin{eqnarray}
{\cal H}_{\rm QSI}
&=&\sum_{<i,j>} \{ J_{zz}S_i^z S_j^z -J_\pm (S_i^+ S_j^- +S_i^-S_j^+)\nonumber \\
        & &+ J_{\pm\pm} [\gamma_{ij}S_i^+S_j^+ + \gamma_{ij}^* S_i^-S_j^-]\nonumber \\
        & &+ J_{z\pm} [(S_i^z(\zeta_{ij}S_j^+ +\zeta_{i,j}^*S_j^-) + i \leftrightarrow j] \} .
\label{Hqsi}
\end{eqnarray}
The parameters for Yb$_2$Ti$_2$O$_7$ determined by fitting 
from high-field inelastic neutron (magnon) spectra in Ref.~\cite{ross} 
are, measured in meV,  
$J_{zz}=0.166\pm 0.04$, $J_\pm=0.05\pm 0.01$, $J_{\pm\pm}=0.05\pm 0.01$,
and $J_{z\pm}=-0.14\pm 0.01$.
Two other sets of parameter estimates for Yb$_2$Ti$_2$O$_7$ were determined by 
fitting the diffused (energy-integrated) neutron scattering using the
random phase approximation (RPA) ~\cite{thompson_prl,chang}.
The values obtained by
Thompson {\it et al.} ~\cite{thompson_prl} are:
 $J_{zz}=0.023$, $J_\pm=0.038$, $J_{\pm\pm}=0.007$,
and $J_{z\pm}=-0.040$, 
while those obtained by Chang {\it et al.} ~\cite{chang} are
 $J_{zz}=0.059$, $J_\pm=0.023$, $J_{\pm\pm}=0.006$,
and $J_{z\pm}=-0.029$. In all cases, the values of the
$\{ J_e \}$ exchange parameters are given in meV. 
 The calculated heat
capacity for all these parameters, together with the experimental data 
on Yb$_2$Ti$_2$O$_7$ from difference groups \cite{ross_prb,yaouanc_prb}, 
are shown in Fig.~S2. It is clear that the latter two
parametrizations by Thompson {\it et al.} and Chang {\it et al.} 
do not give a good description of the heat capacity of the material.
It is not clear at this time why RPA calculations find such
$\{ J_e \}$ parameters compared to high-field paramagnon spectra \cite{ross_prb}.
This problem warrants further attention.

\begin{figure}
\centering
\includegraphics[angle=270,width=10cm]{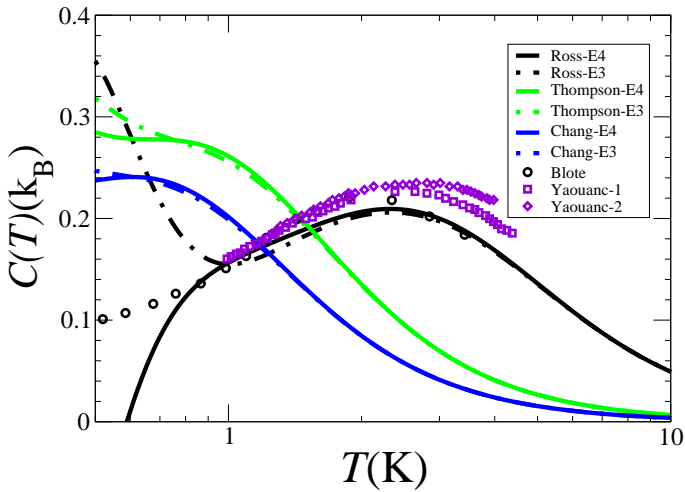}
\caption{S2: Molar heat capacity for YbTO reported by Bl{\" o}te {\it et al.}
~\cite{blote} and by Yaouanc {\it et al.} ~\cite{yaouanc_prb} compared with calculated
values using exchange parameters from Ross {\it et al.} (Ross-E3,E4) ~\cite{ross_prb},
Thompson {\it et al.} (Thompson-E3,E4) ~\cite{thompson_prl} 
and Chang {\it et al.} (Chang-E3,E4) ~\cite{chang}.
Third (E3) and Fourth (E4) order Euler Transforms of the NLC results using the parameters
are  shown. 
}
\label{Cv-YbTO}
\end{figure}

In order to explore to what extent quantum mechanical effects are at
play in ${\cal H}_{\rm QSI}$, we introduce a Hamiltonian with rescaled quantum terms as
\begin{equation}
{\cal H}_\lambda ={\cal H}_0 +\lambda {\cal H}_1,
\end{equation}
where ${\cal H}_0$ is the classical spin-ice Hamiltonian
consisting of $J_{zz}$ terms only, while all other terms
are included in ${\cal H}_1$. The value $\lambda=1$
corresponds to the parameters of Ross {\it et al}.\cite{ross}
In the perturbative regime ($\lambda \ll 1$), this model maps on to a $J_1-J_3$
model with $J_1=J_{zz}$ and $J_3=-3 \lambda^2 J_{z\pm}^2/J_{zz}$.

Specific heat and entropy of the system with different values
of $\lambda$ in 4th order Euler Transform, down to a temperature
where $3$rd and $4$th order Euler Transforms agree with each other are shown in
Fig.~S3 and Fig.~S4. 
Heat capacity of the perturbative classical
$J_1-J_3$ model, calculated by classical loop Monte Carlo 
simulations \cite{loop_MC} is shown in Fig.~S5. Note that
while the models with different $\lambda$
always have a short-range order peak, in the $J_1-J_3$ model,
long-range order temperature increases well past the 
short-range order peak with increasing $J_3/J_1$.

\begin{figure}
\centering
\includegraphics[angle=270,width=10cm]{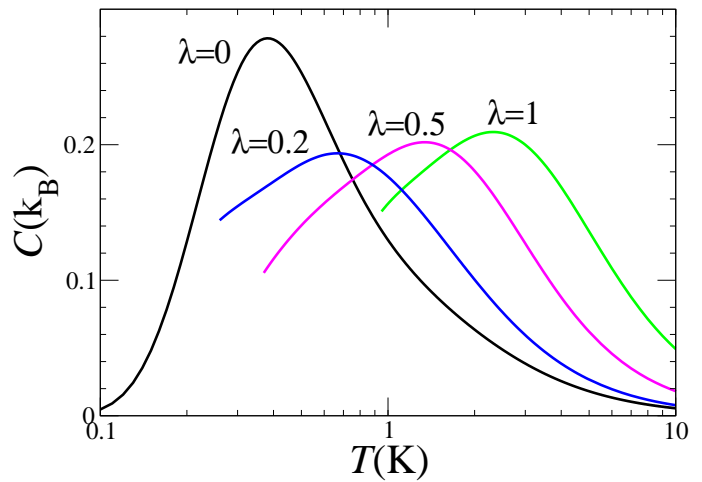}
\caption{S3: Heat capacity where quantum terms are scaled by $\lambda$.
}
\label{Cv-l}
\end{figure}

\begin{figure}
\centering
\includegraphics[angle=270,width=10cm]{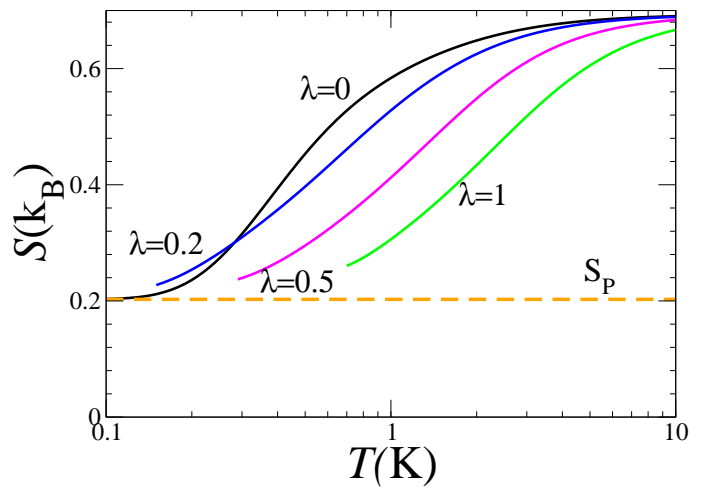}
\caption{S4: Entropy of the system, when quantum terms are scaled
by $\lambda$.
The orange line is the Pauling entropy $S_p$.
}
\label{S-l}
\end{figure}

%%%%%%%%%%%%%%%%%%%%%%%%%%%%%%%%%%%%%%%%%%%%%%%%%%%%%%%%%%%%%%%%%%%%%%%%%%%%%%%%%%%%%%%%%%

\subsection*{Comparison of the experimental entropy vs NLC results}

The entropy difference, $S(T_2) - S(T_1)$ between two temperatures $T_1$ and $T_2$ can 
be obtained by integrating $C(T)/T$ between those two temperatures:
\begin{eqnarray*}
S(T_2) - S(T_1) = \int_{T_1}^{T_2} \frac{C(T)}{T} dT
\end{eqnarray*}

The number of experimental specific heat, $C(T)$, results on Yb$_2$Ti$_2$O$_7$ has rapidly accumulated
over the past year or so \cite{chang,yaouanc_prb,ross_prb}. Most of these data are somewhat
problematic in wanting to assess whether those
thermodynamic data hide spin ice phenomenology, associated with a rapid diminution of
spinon/antispinon excitation and the concurrent $C(T)$ hump at a temperature $\sim 2$ K as we now discuss.

All of the published $C(T)$ data  \cite{chang,blote,yaouanc_prb,ross_prb}
do not go to sufficiently high temperature to extract reliably the limiting $C(T) \propto 1/T^2$ 
high temperature behaviour that would allow one to determine the residual magnetic
entropy by integrating $C(T)/T$ upon decreasing $T$ starting from the infinite $k_{\rm B}\ln(2)$ value.
One must therefore integrate $C(T)/T$ from low temperature, and assume an entropy value, $S_{\rm low}$
at some reference (low) temperature, $T_{\rm low}$.
The apparent large amount of
residual entropy below $\sim 0.2$ K in the single crystal samples of Refs.~\cite{chang,yaouanc_prb,ross_prb}
make difficult ascribing a reasonable value to $S_{\rm low}$. 
This problem is further compounded by the rising low-temperature nuclear contribution 
to the total specific heat below about 0.1 K.
The very sharp 1st order
transition seen in powder powder sample of Ref.~\cite{ross_prb}, without a precise measurement
of the associated latent heat also make difficult using those data for comparison of
experimental entropy with the $S(T)$ calculated by NLC.
On the otherhand, the data of Bl{\" o}te {\it et al.}~\cite{blote} seem the most adequate for
comparison with NLC: there is a sharp specific heat peak at $T_c \sim 0.24$ K with sufficient
temperature resolution that allows integration of $C(T)/T$ over the peak without concern about an
associated latent heat. The $C(T)$ data are dropping rapidly below $T_c$, suggesting the
opening of an excitation gap, ultimately reaching a low-value that is limited by the
``high temperature tail'' ($T \sim 0.1$ K) of the nuclear contribution.
Using the data from Ref.~\cite{blote}, we thus assume that the magnetic part of the
specific heat is zero at $T = 0.1$ K, and integrate {\it upward} (increasing temperature)
$C(T)/T$ up to the highest temperature point available from those data ($\sim 3.5$ K). 
This results in the data (filled black circles in Fig. 3 in the body of the paper).

It would be highly desirable to repeat this procedure from the $C(T)$ data of Refs.~\cite{chang,yaouanc_prb,ross_prb}
which show a sharp peak, but including (magnetic specific heat) 
data for $T$ up to 20 K where the limiting high-temperature regime
$C(T) \approx \frac{A}{T^2} + \frac{B}{T^3}$ can be fitted and compared with NLC, along
with measurements of the magnetic entropy, $S(T)$.

limit 
available dataThe data from 
Working from the reasonable presumption high temperature

%%%%%%%%%%%%%%%%%%%%%%%%%%%%%%%%%%%%%%%%%%%%%%%%%%%%%%%%%%%%%%%%%%%%%%%%%%%%%%%%%%%%%%%%%%

\subsection*{Monte Carlo Simulation of the $J_{zz}-J_3$ Model}

In the perturbative regime of the QSI, we consider the effective Hamiltonian
\begin{equation}
	{\cal H} = \sum_{<i,j>} J_{zz} \sigma_i \sigma_j+ \sum_{<i,j>'} J_3 \sigma_i \sigma_j
	\label{Jzz-J3}
\end{equation}
where $\sigma=\pm 1$ are the Ising variables.
$\langle \ldots  \rangle$ denotes the sum over the nearest neighbors,
$\langle \ldots  \rangle'$ denotes the sum over the third nearest neighbors which share a nearest neighbour.
Distance-wise there exists another type of  third nearest neighbors which do not share a nearest neighbor. 
For any given spin, there are six third nearest neighbors for both types.
Antiferromagnetic $J_{zz}>0$ drives the spin ice formation in the classical spin ice system,
and a small fluctuation-induced ferromagnetic exchange $ J_3\equiv -3J_{z\pm}^2/J_{zz} < 0$  favors 
the ${\bm q}=0$ ordering within the spin ice manifold, {\it i.e.}, 
all tetrahedra on the same primitive FCC lattice have the same one of the six spin ice states.

Monte Carlo simulations are performed using the Metropolis algorithm.
Single spin flip updates are used along with the non-local loop algorithm \cite{loop_MC}, 
which restores the ergodicity of the system once it is frozen into the spin ice states.
Systems of 128 spins are simulated in a cubic box with periodic boundary conditions.
Up to about 78,000 Monte Carlo steps per spin are used in equilibrating the system at a given temperature,
with the same number of steps in data sampling.
To investigate the calorimetric quantities, fluctuations of the energy are recorded 
to give the heat capacity:
\begin{equation}
	C = { < E^2 > - {< E >} ^2 \over k_{\rm B} T^2 }
\end{equation}

\begin{figure}
\centering
\includegraphics[angle=270,width=10cm]{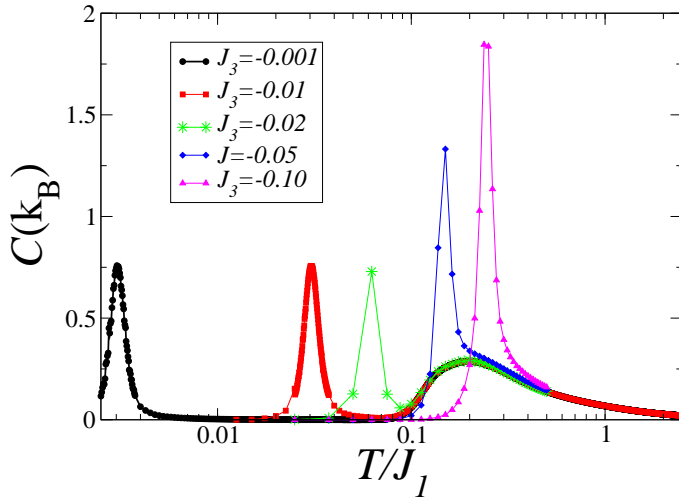}
\caption{S5: Heat capacity of the classical $J_{zz}-J_3$ model,
with different $J_{zz}/J_1$ ratios.
}
\label{Cv-MC}
\end{figure}

%%%%%%%%%%%%%%%%%%%%%%%%%%%%%%%%%%%%%%%%%%%%%%%%%%%%%%%%%%%%%%%%%%%%%%%%%%%%%%%%%%%%%%%%%%

\subsection*{Calculation of Monopole density}

\newcommand{\tr}{\mbox{tr}}

%Z = \tr(e^{-\beta \hat{H}}), %\end{equation}
%which can be computed from the eigenvalue spectrum $\{E_{\alpha}\}$ as
%\begin{equation}
%Z = \sum_{\alpha} e^{-\beta E_{\alpha} }.
%\end{equation}
%The expectation value for the internal energy was computed from the formula 
%\begin{equation}
%\left<E\right> = \tr(\hat{H} e^{-\beta \hat{H}}) / Z.
%\end{equation}
%This quantity was used in turn to compute the entropy as
%\begin{equation}
%\left<S\right>/k_B = \beta \left<E\right> + \log Z,
%\end{equation}
%to which the heat capacity is related as
%\begin{equation}
%\left<C\right> = T \frac{\partial}{\partial T}\left<S\right>.
%\end{equation}

%monopole density
The defect (spinon/antispinon) monopole number $M(T)$,
for a cluster, is evaluated as
\begin{eqnarray}
M(T) & = & \tr(\hat{m} e^{-\beta \hat{H}}) / Z \nonumber \\
 & = & \frac{1}{Z} \sum_{\alpha} e^{-\beta E_{\alpha}} 
\langle\alpha|\hat{m}|\alpha\rangle \nonumber \\
 & = & \frac{1}{Z} \sum_{\alpha, k} e^{-\beta E_{\alpha}} |\langle 
\alpha|k\rangle|^2 m_k
\end{eqnarray}
where $m_k$ is the monopole count in the local $S_z$ basis state $|k 
\rangle$. This count is a sum over all the tetrahedra in a cluster, $m_k = 
\sum_i m_{ki}$, where
\begin{equation}
m_{ki} = \left\{ \begin{array}{rl}
	2 & \text{  all in/out,}\\
	1 & \text{  three in/out and one out/in,}\\
	0 & \text{  two in and two out.}
	\end{array} \right.
\end{equation}
%for the four spins in the $i^{\text{th}}$ tetrahedron.
The monopole density $\rho(T)$ is defined as number of monopoles
present per site, giving
\begin{equation}
\rho(T)=M(T)/N.
\end{equation}

%%%%%%%%%%%%%%%%%%%%%%%%%%%%%%%%%%%%%%%%%%%%%%%%%%%%%%%%%%%%%%%%%%%%%%%%%%%%%%%%%%%%%%%%%%

%%%%%%%%%%%%%%%%%%%%%%%%%%%%%%%%%%%%%%%%%%%%%%%%%%%%%%%%%%%%%%%%%%

%\newpage


\begin{thebibliography}{30}
\bibitem{monopole_vs_spinon} To relate our presentation more directly to the compact lattice 
QED context set in Refs.~\cite{Hermele,Castro-Neto}, in which confinement in three-dimensions
is traditionally referred to the strong (electric charge) coupling, we refrain from
using the language of ``monopoles'' employed in Ref.~\cite{castelnovo} to label local defects
in the ice rule  of the parent classical spin ice.
We use instead the more traditional wording of spinon/antispinon to label finite energy
excitations out of the 2in/2out spin ice manifold.
\bibitem{on_3rd_neighbors} These are geometrically 3rd neighbors on the
pyrochlore lattice but not all 3rd neighbors belong to this category.
\end{thebibliography}
\end{document}